\documentclass[aps,pre,twocolumn,superscriptaddress,showpacs]{revtex4-1}

\usepackage{ulem}
\usepackage{graphicx}
\usepackage{dcolumn}
\usepackage{bm}
\usepackage{accents}
\usepackage{color}
\newcommand{\beq}{\begin{equation}}
\newcommand{\eeq}{\end{equation}}
\newcommand{\beqa}{\begin{eqnarray}}
\newcommand{\eeqa}{\end{eqnarray}}
\newcommand{\bew}{\begin{widetext}}
\newcommand{\eew}{\end{widetext}}
\newcommand{\Tr}{\text{Tr}}

\newcommand{\av}[1]{\left\langle #1 \right\rangle}

\begin{document}


\title{Path integral approach to heat in quantum thermodynamics}


\author{Ken Funo}
\affiliation{School of Physics, Peking University, Beijing 100871, China}
\author{H. T. Quan}
\email{htquan@pku.edu.cn}
\affiliation{School of Physics, Peking University, Beijing 100871, China}
\affiliation{Collaborative Innovation Center of Quantum Matter, Beijing 100871, China}
\date{\today}

\date{\today}

\begin{abstract}
We study the heat statistics of a quantum Brownian motion described by the Caldeira-Leggett model. By using the path integral approach, we introduce a novel concept of the quantum heat functional along every pair of Feynman paths. This approach has an advantage of improving our understanding about heat in quantum systems. First, we demonstrate the microscopic reversibility of the system by connecting the heat functional to the forward and its time-reversed probabilities. Second, we analytically prove the quantum-classical correspondence of the heat functional and their statistics, which allows us to obtain better intuitions about the difference between classical and quantum heat.
\end{abstract}


\maketitle

\section{Introduction}

In recent years, techniques for controlling various quantum systems have been put forward. Also, measurement techniques for the thermal and quantum fluctuations in small quantum systems have been significantly improved. It is now at the stage of using those techniques for the experimental studies of the nonequilibrium statistical mechanics ~\cite{Pekola,quantumjarexp}. These attempts are not just limited to verifying the fundamental relations in nonequilibrium statistical mechanics known as fluctuation theorems and the Jarzynski equality, but also as a starting point to apply nonequilibrium statistical mechanics for designing efficient heat transferring quantum devices. Therefore, quantum thermodynamics, which has been studied since 1950s~\cite{1959}, has received renewed interest as the frontier of nonequilibrium statistical mechanics, quantum information theory and nanoscopic physics~\cite{Strasberg,fluctuation1,Esposito,Hanggiaspects,Anders,Horodecki,Parrondo}. 

Our understandings about work and heat in small systems have been significantly improved in the past two decades, known as classical stochastic thermodynamics~\cite{Seifert,Sekimoto,Crooks,Jarzynski1}. On the contrary, the framework of quantum stochastic thermodynamics~\cite{Tasaki,Kurchan,Esposito,fluctuation1,Hanngi09,Horowitz1,Hekking} is far from being well established. In the weak-coupling, Markovian and rotating-wave approximation (RWA) regime, a framework based on the quantum jump method has been established~\cite{Horowitz1,Hekking}. However, a connection to the well established framework of classical stochastic thermodynamics is lacking, and how to identify the genuine quantum effect in quantum thermodynamics is not straightforward. Also, there are limitations on the time variation of the Hamiltonian of the system, i.e., time variation should be either periodic, near adiabatic, or treated as a weak perturbation~\cite{Feireview}.  In Ref.~\cite{Funo17}, we have introduced the work functional along individual Feynman path and studied the work statistics by using the path integral methods for a quantum Brownian motion model~\cite{CaldeiraBook,Hanggi05C}.  We note that this model is applicable to strong-coupling, non-Markovian and non-RWA regime, and the time variation of the Hamiltonian can be arbitrary. Moreover, this approach has an advantage of connecting the classical and quantum work through $\hbar$ expansions.

In this paper, we extend the path integral approach developed in Ref.~\cite{Funo17} and study the heat statistics in a quantum Brownian motion model. Relevant studies can be found in Refs.~\cite{Aurell,Aurell17,Kato,Saito}. In order to avoid subtleties about defining the heat in the strong-coupling regime~\cite{Talkner16}, we limit ourselves to the weak-coupling case. We would like to point out that the usual settings in the classical stochastic thermodynamics~\cite{Seifert,Sekimoto} assumes weak-coupling (see Ref.~\cite{Jarzynski17,Seifert16,Talkner16} for the strong-coupling case), and thus it is natural to consider the same regime in studying the quantum heat as a first step. From the previous studies for the path integral expression for the heat statistics~\cite{Carrega15,Carrega16,Ueda}, we introduce the quantum heat functional along every pair of Feynman paths, and study its properties. In particular,  we show the microscopic reversibility~(\ref{DBR}) using the heat functional. This result connects the heat functional and the probabilities of the forward and the time-reversed paths, and is the most fundamental principle underlying various types of the fluctuation theorem. We further show the quantum-classical correspondence of the quantum heat functional and its statistics.

This article is organized as follows.  In Sec.~\ref{sec:setup}, we briefly explain the setup of the paper by introducing the Caldeira-Leggett model, and introduce the heat and its statistics based on two-point measurement. In Sec.~\ref{sec:heat}, we introduce the heat functional along every pair of Feynman paths and its expansion. In Sec.~\ref{sec:microscopic}, we show the microscopic reversibility and Jarzynski's equality. In Sec.~\ref{sec:classical}, we take the classical limit of the heat functional and prove the quantum-classical correspondence of the heat functional and its statistics. We summarize our result in Sec~\ref{sec:summary}. 

\section{\label{sec:setup}Heat and heat statistics in quantum Brownian motion}

\subsection{Caldeira-Leggett model}
We consider the quantum Brownian motion described by the Caldeira-Leggett model~\cite{CaldeiraBook}, where the Hamiltonian of the composite system is given by
 $H_{\text{tot}}(\lambda_{t})=H_{\rm{S}}(\lambda_{t})+H_{\rm{B}}+H_{\rm{SB}}$, with
\beqa
& &H_{\rm{S}}(\lambda_{t})=\frac{\hat{p}^{2}}{2M}+\hat{V}(\lambda_{t},\hat{x}), H_{\rm{B}}=\sum_{k}\left(\frac{\hat{p}_{k}^{2}}{2m_{k}}+\frac{m_{k}\omega_{k}^{2}}{2}\hat{q}_{k}^{2}\right), \nonumber \\
& &H_{\rm{SB}}=-\hat{x}\otimes\sum_{k}c_{k}\hat{q}_{k}+\sum_{k}\frac{c^{2}_{k}}{2m_{k}\omega_{k}^{2}}\hat{x}^{2}, \label{Hamiltonian}
\eeqa
where $p$, $M$, and $x$ denote the momentum, the mass and the position of the particle; $p_{k}$, $m_{k}$, and $x_{k}$ denote the momentum, the mass and the position of the $k$-th degree of freedom of the heat bath; $\lambda_{t}$ and $V(\lambda_{t},x)$ denote the externally controlled work parameter and the potential of the particle; $\omega_{k}$ and $c_{k}$ denote the frequency of and the coupling strength to the $k$-th degree of freedom of the heat bath. Because we assume a linear coupling between the system and the heat bath in $H_{\rm{SB}}$, we can analytically trace out the bath degrees of freedom, which brings important insights to the understanding of heat in open quantum systems. The classical limit of this model reproduces the Langevin equation, which is a prototypical model used in stochastic thermodynamics.

We assume a weak-coupling between the system and the bath and thus consider the following initial state
\beq
\rho(0)=\rho_{\rm{S}}(0)\otimes \rho_{\rm{B}}^{\rm{G}}. \label{initialrho}
\eeq
Here, $\rho_{\rm{S}}(0)$ is an arbitrary state of the system, $\rho_{\rm{B}}^{\rm{G}}=e^{-\beta H_{\rm{B}}}/Z_{\rm{B}}$, $\beta$ and $Z_{\rm{B}}$ are the Gibbs distribution, inverse temperature and the partition function of the bath, respectively. The reduced density matrix of the system at time $\tau$ is given by $\rho_{\rm{S}}(\tau)=\Tr_{B}[U_{\rm{SB}}\rho(0)U^{\dagger}_{\rm{SB}}]$, where $U_{\rm{SB}}=\hat{\text{T}}[\exp(-\frac{i}{\hbar}\int^{\tau}_{0}dt H_{\text{tot}}(\lambda_{t}))]$ is the unitary time-evolution operator for the composite system. The path integral expression of $\rho_{\rm{S}}(\tau)$ takes the form~\cite{Caldeira83,CaldeiraBook,Feynman,Weiss}
\beqa
\langle x_{f}|\rho_{\rm{S}}(\tau)| y_{f}\rangle &=& \int dx_{i}dy_{i}\int^{x(\tau)=x_{f}}_{x(0)=x_{i}} Dx \int^{y(\tau)=y_{f}}_{y(0)=y_{i}} Dy  \nonumber \\
& &\times e^{\frac{i}{\hbar}(S[x]-S[y])}F_{\text{FV}}[x,y]\rho_{\rm{S}}(x_{i},y_{i}) , \label{rho}
\eeqa
where
\beqa
F_{\text{FV}}[x,y]:=\exp\biggl[& &+\frac{1}{\hbar}\int^{\tau}_{0}dt\int^{\tau}_{0}ds L(s-t)x(t)y(s)\nonumber \\
& & -\frac{1}{\hbar}\int^{\tau}_{0}dt\int^{t}_{0}ds L(t-s)x(t)x(s) \nonumber\\
& &-\frac{1}{\hbar}\int^{\tau}_{0}dt\int^{t}_{0}ds L^{*}(t-s)y(t)y(s) \nonumber \\
& & -\frac{i\mu}{\hbar}\int^{\tau}_{0} dt (x^{2}(t)-y^{2}(t)) \biggr] \label{FV}
\eeqa
is the Feynman-Vernon influence functional~\cite{Feynman},
\beq
L(t):=\sum_{k}\frac{c_{k}^{2}}{2m_{k}\omega_{k}}\Bigl( \coth\frac{\hbar\beta\omega_{k}}{2}\cos\omega_{k} t-i\sin\omega_{k} t\Bigr) \label{bathcorr}
\eeq
is the complex bath correlation function, $\mu:=\sum_{k}c^{2}_{k}/(2m_{k}\omega_{k}^{2})$, and $x(t)$, $y(t)$ are the forward and backward coordinates.  In Eq.~(\ref{rho}), we use the position representation of the initial density matrix $\rho_{\rm{S}}(x_{i},y_{i}):=\langle x_{i}|\rho_{\rm{S}}(0)|y_{i}\rangle$ and the action of the system $S[x]=\int^{\tau}_{0}dt \mathcal{L}[\lambda_{t},x(t)]$, where $\mathcal{L}[\lambda_{t},x(t)]:=\frac{M}{2}\dot{x}^{2}(t)-V(\lambda_{t},x(t))$ is the Lagrangian. We note that the usual path integral expression for a pure state requires only a single Feynman path $x(t)$. However, for a mixed state, we need a pair of Feynman paths $\{x(t),y(s)\}$ to describe the time evolution as shown in Eq.~(\ref{rho}). 

For later convenience, we define the path probability of the system conditioned on $(x_{i},y_{i})$ as follows:
\beq
\mathcal{P}[x(t),y(s)|x_{i},y_{i}]:=e^{\frac{i}{\hbar}(S[x]-S[y])}F_{\text{FV}}[x,y]. \label{Forwardprob}
\eeq
Here, the boundary conditions for the forward and backward coordinates are given by $x(0)=x_{i}$, $y(0)=y_{i}$, $x(\tau)=x_{f}$ and $y(\tau)=y_{f}$. We note that Eq.~(\ref{Forwardprob}) satisfies the usual normalization condition for the conditional probability:
\beq
\int dx_{f}dy_{f}\delta(x_{f}-y_{f})\int Dx Dy\mathcal{P}[x(t),y(s)|x_{i},y_{i}]=1.
\eeq
Using this path probability~(\ref{Forwardprob}), the reduced density matrix of the system at time $\tau$ can be obtained as follows:
\beqa
& &\langle x_{f}|\rho_{\rm{S}}(\tau)|y_{f}\rangle \nonumber \\
&=&\int dx_{i}dy_{i}\int DxDy \mathcal{P}[x(t),y(s)|x_{i},y_{i}]\rho_{\rm{S}}(x_{i},y_{i}).
\eeqa


\subsection{Two-point measurement based definition of heat functional and its statistics}
We adopt the two-point measurement definition and measure the energy of the bath twice at time $t=0$ and $t=\tau$ and obtain $E_{\rm{B}}(m)$ and $E_{\rm{B}}(m')$, respectively. In the weak coupling regime, we can neglect the interaction energy when defining the heat. We thus define the fluctuating heat as the difference in the measured energies of the bath:
\beq
Q_{m,m'}:=E_{\rm{B}}(m)-E_{\rm{B}}(m').
\eeq
Here, the probability of obtaining the outcomes $m$ and $m'$ is given by
\beq
p(m,m'):=\text{Tr}_{\rm{SB}}[ P_{\rm{B}}^{m'} U_{\rm{SB}} (\rho_{\rm{S}}(0)\otimes P_{\rm{B}}^{m}) U_{\rm{SB}}^{\dagger} ]\frac{e^{-\beta E_{\rm{B}}(m)}}{Z_{\rm{B}}},
\eeq
where $P_{\rm{B}}^{m}:=|E_{\rm{B}}(m)\rangle\langle E_{\rm{B}}(m)|$ is the projection operator using the energy eigenstate $|E_{\rm{B}}(m)\rangle$. The heat probability distribution $P(Q)$ is then defined as
\beq
P(Q):=\sum_{m,m'}\delta(Q-Q_{m,m'})p(m,m').
\eeq
 By taking the Fourier transformation of $P(Q)$, we obtain the characteristic function of heat $\chi_{Q}(\nu)$ as follows:
\beq
\chi_{Q}(\nu)=\Tr\Bigl[ U_{\rm{SB}}e^{i\nu H_{\rm{B}}}(\rho_{\rm{S}}(0)\otimes \rho_{\rm{B}}^{\rm{G}} ) U^{\dagger}_{\rm{SB}}e^{-i\nu H_{\rm{B}}}\Bigr]. \label{supp:EXHB}
\eeq
From Eq.~(\ref{supp:EXHB}), the $n$-th moment of the heat distribution can be obtained from the formula
\beq
\langle Q^{n}\rangle= \left. (-i)^{n}\partial^{n}_{\nu}\chi_{Q}(\nu)\right|_{\nu=0}. \label{heatnformula}
\eeq



\section{\label{sec:heat}Feynman path based definition of heat functional and its statistics}
In this section, we use the path integral formalism and rewrite the characteristic function of heat. Though the characteristic function of heat based on Feynman path~(\ref{CFH}) is equivalent to that based on the two-point measurement~(\ref{supp:EXHB}), we will show in the following that the Feynman path-based definition of heat functional brings more insights to the understanding of heat in open quantum systems.

We now use the path integral technique and integrate out the bath degrees of freedom in Eq.~(\ref{supp:EXHB}) and express $\chi_{Q}(\nu)$ by using only the degrees of freedom of the system. 
 The path integral expression for the characteristic function of heat is now given by~\cite{Carrega15,Ueda} (see Appendix~\ref{supp:sectwo} for the derivation)
\beqa
\chi_{Q}(\nu)&=&\int dx_{f}dy_{f}dx_{i}dy_{i}\delta(x_{f}-y_{f}) \int Dx\int Dy \label{CFH} \\
& &\times e^{\frac{i}{\hbar}(S[x]-S[y])}F_{\text{FV}}\left[x,y\right]\rho_{\rm{S}}(x_{i},y_{i})e^{i\nu Q_{\nu}[x,y]}. \nonumber
\eeqa
Here, the heat functional $Q_{\nu}[x,y]$ is defined along every pair of Feynman paths $\{x(t),y(s)\}$ as
\beqa
& &Q_{\nu}[x,y] \label{Qfunction} \\
&:=&\frac{-i}{\hbar\nu}\int^{\tau}_{0}dt\int^{\tau}_{0}ds \Bigl( L(s-t+\hbar\nu)-L(s-t)\Bigr)x(t)y(s). \nonumber
\eeqa

  Using the formula~(\ref{heatnformula}), the path integral expression for the $n$-th moment of the heat distribution is given by
\beq
\langle Q^{n}\rangle=\int \left. \mathcal{P}[x,y|x_{i},y_{i}]\rho_{\rm{S}}(x_{i},y_{i}) (-i)^{n}\partial^{n}_{\nu}  e^{i\nu Q_{\nu}[x,y]}\right|_{\nu=0}, \label{heatnpath}
\eeq
where the integration should be done over $\int dx_{f}dy_{f}dx_{i}dy_{i}\delta(x_{f}-y_{f}) \int Dx\int Dy$. When applying the formula~(\ref{heatnpath}), it is convenient to expand $i\nu Q_{\nu}[x,y]$ in terms of $\nu$ as follows:
\beq
i\nu Q_{\nu}[x,y]=\sum_{n=1}\frac{(i\nu)^{n}}{n!}Q^{(n)}[x,y],
\eeq
with
\beq
Q^{(n)}[x,y]:=(-i)^{n}\hbar^{n-1}\hspace{-0.3mm}\int^{\tau}_{0}\hspace{-0.6mm}dt\hspace{-0.1mm}\int^{\tau}_{0}\hspace{-0.6mm}ds\hspace{-0.1mm} \left(\partial_{s}^{n}L(s-t)\right)x(t)y(s).
\eeq
We then find that the first moment of the heat distribution is given by
\beq
\langle Q\rangle=\int \mathcal{P}[x,y|x_{i},y_{i}]\rho_{\rm{S}}(x_{i},y_{i})Q^{(1)}[x,y]. \label{firstHmoment}
\eeq
However, from the noncommutativity in quantum mechanics, the second moment of the heat distribution deviates from the path-integral average of $(Q^{(1)}[x,y])^{2}$ as follows:
\beqa
& &\langle Q^{2}\rangle  \label{secondHmoment} \\
&=&\int \mathcal{P}[x,y|x_{i},y_{i}]\rho_{\rm{S}}(x_{i},y_{i}) \left\{ \left(Q^{(1)}[x,y]\right)^{2}+Q^{(2)}[x,y] \right\}. \nonumber
\eeqa
We would like to emphasize that the heat functional is a trajectory-dependent quantity, and it cannot be expressed as an observable of a single operator~\cite{characteristic1}. Therefore, the second term on the right-hand side of Eq.~(\ref{secondHmoment}) describes the effect of the noncommutativity between operators. Similarly, $\langle Q^{n}\rangle$ is not equal to the path integral average of $(Q^{(1)}[x,y])^{n}$. There are additional terms due to the noncommutativity, and they are contained in the $\nu$-dependence of $Q_{\nu}[x,y]$.




By taking $\nu=-i\beta$, we have
\beq
Q_{\beta}[x,y]=\frac{1}{\hbar\beta}\int^{\tau}_{0}dt\int^{\tau}_{0}ds \Bigl( L(t-s)-L(s-t)\Bigr)x(t)y(s).\label{betaHeat}
\eeq
Here, we use the property of the bath correlation function: $L(t-i\beta)=L^{*}(t)=L(-t)$. Because the heat functional is related to the time-asymmetric part of the two-point bath correlation function, Eq.~(\ref{betaHeat}) is related to the ratio of the forward and the time-reversed path probabilities through the microscopic reversibility as we will show in Eq.~(\ref{DBR}).

\section{\label{sec:microscopic}Microscopic reversibility and Jarzynski's equality in quantum Brownian motion model}
We now show our first main result, the microscopic reversibility of the system by utilizing the heat functional introduced in Eq.~(\ref{betaHeat}). We then utilize the microscopic reversibility relation and show the path integral derivation of Jarzynski's equality.

\subsection{\label{sec:detailed}Microscopic reversibility}
We now consider the probability of the time-reversed path. We define the time-reversed time $\tilde{t}:=\tau-t$ and the time-reversed position of the system $\tilde{x}(\tilde{t}):=x(t)$.

Now the time-reversal of the Feynman-Vernon influence functional can be related to the heat functional as follows:
\beqa
\tilde{F}_{\text{FV}}[\tilde{x},\tilde{y}]&:=&\exp\biggl[ +\frac{1}{\hbar}\int^{\tau}_{0}d\tilde{t}\int^{\tau}_{0}d\tilde{s} L(\tilde{s}-\tilde{t})\tilde{x}(\tilde{t})\tilde{y}(\tilde{s})  \nonumber \\
& &\ \ \ \ \ \  -\frac{1}{\hbar}\int^{\tau}_{0}d\tilde{t}\int^{\tilde{t}}_{0}d\tilde{s} L(\tilde{t}-\tilde{s})\tilde{x}(\tilde{t})\tilde{x}(\tilde{s}) \nonumber\\
& &\ \ \ \ \ \ -\frac{1}{\hbar}\int^{\tau}_{0}d\tilde{t}\int^{\tilde{t}}_{0}d\tilde{s} L^{*}(\tilde{t}-\tilde{s})\tilde{y}(\tilde{t})\tilde{y}(\tilde{s}) \nonumber \\
& &\ \ \ \  \ \ -\frac{i\mu}{\hbar}\int^{\tau}_{0} d\tilde{t} (\tilde{x}^{2}(\tilde{t})-\tilde{y}^{2}(\tilde{t}))\biggr] \nonumber \\
&=&\exp\biggl[ +\frac{1}{\hbar}\int^{\tau}_{0}dt\int^{\tau}_{0}ds L(t-s)x(t)y(s) \nonumber \\
& &\ \ \ \  \ \ -\frac{1}{\hbar}\int^{\tau}_{0}dt\int^{t}_{0}ds L(t-s)x(t)x(s) \nonumber\\
& &\ \ \ \ \ \ -\frac{1}{\hbar}\int^{\tau}_{0}dt\int^{t}_{0}ds L^{*}(t-s)y(t)y(s) \nonumber \\
& &\ \ \ \ \ \  -\frac{i\mu}{\hbar}\int^{\tau}_{0} dt (x^{2}(t)-y^{2}(t)) \biggr] \nonumber \\
&=&F_{\text{FV}}[x,y]\exp(\beta Q_{\beta}[x,y]). \label{FVt}
\eeqa
Note that the last three terms inside the exponent of $F_{\rm{FV}}[x,y]$~(\ref{FV}) are time-symmetric, but the first term
\beq
\hbar^{-1}\int^{\tau}_{0}dt \int^{\tau}_{0}ds L(s-t)x(t)y(s) \label{FVassym}
\eeq
is time-asymmetric. Since the difference between Eq.~(\ref{FVassym}) and its time-reversed one is found to be equal to $\beta Q_{\beta}[x,y]$ (\ref{betaHeat}), we obtain the last equality in Eq.~(\ref{FVt}).

Similarly, the time-reversed action is given by
\beqa
\tilde{S}[\tilde{x}]&:=&-\int^{\tau}_{0}d\tilde{t}\Bigl( \frac{M}{2}\dot{\tilde{x}}^{2}(\tilde{t})-V(\tilde{\lambda}_{\tilde{t}},\tilde{x}(\tilde{t}))\Bigr) \nonumber \\
&=&S[x].
\eeqa
Here, $\tilde{\lambda}_{\tilde{t}}:=\lambda_{\tau-\tilde{t}}$. By defining the time-reversed path probability conditioned on $(x_{f},y_{f})$ as
\beq
\tilde{\mathcal{P}}[\tilde{x}(\tilde{t}),\tilde{y}(\tilde{s})|x_{f},y_{f}]:=e^{\frac{i}{\hbar}(\tilde{S}[\tilde{x}]-\tilde{S}[\tilde{y}])}\tilde{F}_{\text{FV}}[\tilde{x},\tilde{y}],
\eeq
we obtain the following microscopic reversibility relation (detailed balance relation)
\beq
\frac{\tilde{\mathcal{P}}[\tilde{x}(\tilde{t}),\tilde{y}(\tilde{s})|x_{f},y_{f}]}{\mathcal{P}[x(t),y(s)|x_{i},y_{i}]}=e^{\beta Q_{\beta}[x,y]} , \label{DBR}
\eeq
connecting the ratio of the conditional forward and time-reversed probabilities and the heat functional. Therefore, we have shown that the heat functional~(\ref{betaHeat}) is related to the time-reversal symmetry of the system along every pair of Feynman paths $\{x(t),y(s)\}$. For isolated systems, $\tilde{F}_{\text{FV}}[\tilde{x},\tilde{y}]=F_{\text{FV}}[x,y]=1$, so we have $Q_{\beta}[x,y]=0$, which agrees with our intuitions.

We note that the microscopic reversibility of the system~(\ref{DBR}) is the most fundamental relation in deriving various types of fluctuation theorems~\cite{Crooks,Seifert}. We also note that this microscopic reversibility relation has been obtained for open quantum systems described by the Lindblad master equation \cite{Horowitz1}. But for the quantum Brownian motion model, it has not been shown so far. Thus we extend the most fundamental fluctuation theorem to quantum Brownian motion model.

\begin{figure}[t]
\begin{center}
\includegraphics[width=.35\textwidth]{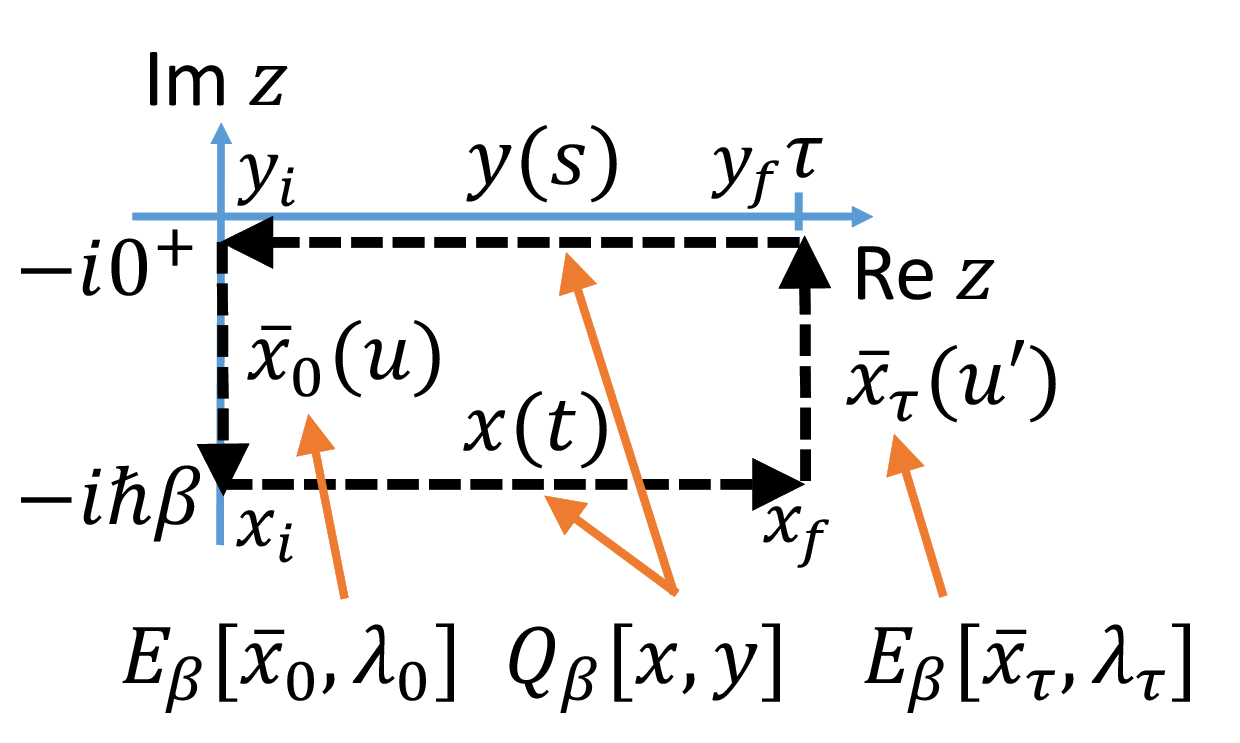}
\caption{Contour used in the path integral formalism to define the internal energy functional and the heat functional. Here, $\bar{x}_{0}(u)$ is the imaginary time coordinate of the system at $t=0$, which is used to define the initial energy functional $E_{\beta}[\bar{x}_{0},\lambda_{0}]$~(\ref{initialEfunctional}). Similarly, we define the final energy functional $E_{\beta}[\bar{x}_{\tau},\lambda_{\tau}]$ by using $\bar{x}_{\tau}(u')$. The heat functional $Q_{\beta}[x,y]$~(\ref{betaHeat}) is defined as the time-asymmetric part of the two-point bath correlation function connecting the forward $x(t)$ and backward $y(s)$ coordinates.
}
\label{fig:one}
\end{center}
\end{figure}

\subsection{Path integral derivation of Jarzynski's equality}
Having established a connection between the heat functional and the ratio of the conditional forward and time-reversed path probabilities~(\ref{DBR}), we derive Jarzynski's equality based on path integral formalism. See also Refs.~\cite{Hu,Funo17} for the derivation of Jarzynski's equality using different methods. Let us assume that the initial state of the system is given by the canonical distribution, i.e., $\rho_{\rm{S}}(0)=\exp(-\beta H_{\rm{S}}(\lambda_{0}))/Z_{\rm{S}}(\lambda_{0})$. Then, we can use the imaginary time integral and express the matrix element of the initial density matrix as
\beq
\langle x_{i}|\rho_{\rm{S}}(0)|y_{i}\rangle=\frac{1}{Z_{\rm{S}}(\lambda_{0})}\int^{\bar{x}(\hbar\beta)=x_{i}}_{\bar{x}(0)=y_{i}}D\bar{x}_{0} e^{-\frac{1}{\hbar}S^{\mathrm{E}}[\bar{x}_{0},\lambda_{0}]},
\eeq
where
\beq
S^{\mathrm{E}}[\bar{x}_{0},\lambda_{0}]:=\int^{\hbar\beta}_{0}du \left(\frac{M}{2}[\dot{\bar{x}}_{0}(u)]^{2}+V(\lambda_{0},\bar{x}_{0}(u))\right)
\eeq
is the Euclidian action, and $\bar{x}_{0}$ is the imaginary time-coordinate of the system.

We now define the initial energy functional of the system as
\beq
E_{\beta}[\bar{x}_{0},\lambda_{0}]:=\frac{1}{\hbar\beta}S^{\mathrm{E}}[\bar{x}_{0},\lambda_{0}]. \label{initialEfunctional}
\eeq
We note that Eq.~(\ref{initialEfunctional}) reproduces the initial energy statistics of the system, as shown in Appendix~\ref{supp:secE}. Similar to Eq.~(\ref{initialEfunctional}), we introduce another imaginary time-coordinate $\bar{x}_{\tau}$ connecting $x_{f}$ and $y_{f}$ (see Fig.~\ref{fig:one}), and define the final energy functional of the system as $E_{\beta}[\bar{x}_{\tau},\lambda_{\tau}]:=\frac{1}{\hbar\beta}S^{\mathrm{E}}[\bar{x}_{\tau},\lambda_{\tau}]$. Now the energy difference of the system is defined as
\beq
\Delta E_{\beta}[\bar{x}_{0},\bar{x}_{\tau}]:=E_{\beta}[\bar{x}_{\tau},\lambda_{\tau}]-E_{\beta}[\bar{x}_{0},\lambda_{0}],
\eeq
and we define the work functional by utilizing the first law of thermodynamics:
\beq
W_{\beta}[\bar{x}_{0},x,y,\bar{x}_{\tau}]:=\Delta E_{\beta}[\bar{x}_{0},\bar{x}_{\tau}]-Q_{\beta}[x,y]. \label{Wfunctional}
\eeq
By using Eq.~(\ref{Wfunctional}), we calculate the path integral average of the exponentiated work (see Fig.~\ref{fig:one} for the contour we use for the path integral)
\beqa
& &\av{e^{-\beta W_{\beta}}}:=\int dx_{f}dy_{f}dx_{i}dy_{i}\int D\bar{x}_{\tau}DxDyD\bar{x}_{0}  \\
& &\ \ \ \ \times \mathcal{P}[x(t),y(s)|x_{i},y_{i}]\frac{e^{-\frac{1}{\hbar}S^{\mathrm{E}}[\bar{x}_{0},\lambda_{0}]}}{Z_{\rm{S}}(\lambda_{0})}e^{-\beta W_{\beta}[\bar{x}_{0},x,y,\bar{x}_{\tau}]}. \nonumber
\eeqa
By using the microscopic reversibility~(\ref{DBR}) and Eq.~(\ref{Wfunctional}), we have
\beqa
\av{e^{-\beta W_{\beta}}}&=&\int dx_{f}dy_{f}dx_{i}dy_{i}\int D\bar{x}_{\tau}DxDyD\bar{x}_{0} \nonumber  \\
& &\times \tilde{\mathcal{P}}[\tilde{x}(\tilde{t}),\tilde{y}(\tilde{s})|x_{f},y_{f}]\frac{e^{-\frac{1}{\hbar}S^{\mathrm{E}}[\bar{x}_{\tau},\lambda_{\tau}]}}{Z_{\rm{S}}(\lambda_{0})} \nonumber \\
&=&\frac{Z_{\rm{S}}(\lambda_{\tau})}{Z_{\rm{S}}(\lambda_{0})}, \label{proofJE}
\eeqa
where we have used the normalization condition of the time-reversed path probability
\beqa
1&=&\int dx_{f}dy_{f}dx_{i}dy_{i}\int D\bar{x}_{\tau}DxDyD\bar{x}_{0}   \nonumber \\
& &\times \tilde{\mathcal{P}}[\tilde{x}(\tilde{t}),\tilde{y}(\tilde{s})|x_{f},y_{f}]\frac{e^{-\frac{1}{\hbar}S^{\mathrm{E}}[\bar{x}_{\tau},\lambda_{\tau}]}}{Z_{\rm{S}}(\lambda_{\tau})}
\eeqa
and derive the last line in Eq.~(\ref{proofJE}). Therefore, we derive Jarzynski's equality based on the microscopic reversibility of the system~(\ref{DBR}):
\beq
\langle e^{-\beta (W_{\beta}-\Delta F_{\rm{S}})} \rangle=1.
\eeq

\section{\label{sec:classical}Quantum classical correspondence of the heat functional and its statistics}
We show our second main result in this section by proving the quantum-classical correspondence of the heat statistics for the Caldeira-Leggett model.

Let us first consider the classical limit of the heat functional by expanding Eq.~(\ref{Qfunction}) in terms of $\hbar\nu$:
\beq
 Q_{\nu}[x,y] =-i\int^{\tau}_{0}dt\int^{\tau}_{0}ds \dot{L}(s-t)x(t)y(s)+O(\hbar\nu) . \label{Qlimita}
\eeq
We now use the notation $X(t)=(x(t)+y(t))/2$ and $\xi(t)=x(t)-y(t)$. We note that when $\hbar \to 0$, we have $\xi=O(\hbar)$ and $X = x+O(\hbar)=y+O(\hbar)$. Therefore, $X$ behaves as the classical trajectory of the system and $\xi$ describes quantum fluctuations. We now introduce the noise function
\beq
\Omega(t):=i\int^{\tau}_{0}ds L_{\text{Re}}(t-s)\xi(s) , \label{supp:NOISE}
\eeq
with $L_{\text{Re}}(t):=\text{Re}[L(t)]$, satisfying the following properties:
\beqa
& &\av{\Omega(t)}=0,\\
& &\av{\Omega(t)\Omega(s)}=\hbar L_{\text{Re}}(t-s)=\beta^{-1}K(t-s)+O(\beta), \label{noisecorre} \hspace{5mm}
\eeqa
where the average over the noise is defined by $\av{f[\Omega(t)]}:=\int D\Omega P[\Omega]f[\Omega(t)]$ and
\beq
P[\Omega]:=C^{-1}\exp\bigl[ -\frac{1}{2\hbar}\int^{\tau}_{0}dt\int^{\tau}_{0}ds \Omega(t)L^{-1}_{\text{Re}}(t-s)\Omega(s)\Bigr], \label{weightF}
\eeq
is the weight function with the normalization constant $C$. In Eq.~(\ref{noisecorre}), we use the classical bath correlation function
\beq
K(t-s):=\sum_{k}\frac{c_{k}^{2}}{m_{k}\omega_{k}^{2}}\cos\omega_{k}(t-s). \label{clbathcorre}
\eeq
Therefore, in the high temperature limit $\beta\rightarrow 0$, $\Omega(t)$ satisfies the property of the classical noise~(\ref{noisecorre}). 

Using the noise functional~(\ref{supp:NOISE}) and the classical trajectory of the system $X(t)$, we finally obtain (see Appendix~\ref{appendix:cal} for details)
\beq
Q_{\nu}[x,y]=Q_{\text{cl}}[X,\Omega]+Q_{\text{int}}[X,\Omega]+Q_{\text{slip}}[X,\Omega]+O(\hbar), \label{heatdecomp}
\eeq
where
\beq
Q_{\text{cl}}[X,\Omega]=\int^{\tau}_{0}\hspace{-1mm}dt \dot{X}(t)\Bigl( \Omega(t)-\int^{t}_{0}\hspace{-1mm}ds K(t-s)\dot{X}(s) \Bigr) \label{supp:classicalq}
\eeq
is the classical trajectory heat for the non-Markovian dynamics [Eq.~(\ref{supp:CLQU})],
\beqa
Q_{\text{int}}[X,\Omega] &:=&X(0)X(\tau)K(\tau)-\frac{1}{2}\Bigl(X^{2}(0)+X^{2}(\tau)\Bigr)K(0) \nonumber \\
& & +\frac{1}{2}X(\tau)\int^{\tau}_{0}dt\dot{X}(t)K(\tau-t)  \nonumber \\
& &-X(\tau)\Omega(\tau)+X(0)\Omega(0) ,
 \label{supp:qbd}
\eeqa
is equal to the change of the interaction energy $\Delta H_{\rm{SB}}$ for the classical Brownian motion model [Eq.~(\ref{supp:deltaSB})], and
\beq
Q_{\text{slip}}[X,\Omega]:=-X(0)\int^{\tau}_{0}dtK(t)\dot{X}(t) \label{slip}
\eeq
is the heat generated by a fast relaxation (initial slippage) of the bath to the conditional canonical distribution for a given initial state $X(0)$~\cite{Carrega15}. In the weak coupling regime, we can neglect the interaction energy $\Delta H_{\rm{SB}}$ and also the initial slippage effect $Q_{\text{slip}}[X,\Omega]$ (see Appendix~\ref{supp:secone} for details). Therefore we have 
\beq
Q_{\nu}[x,y]=Q_{\text{cl}}[X,\Omega]+O(\hbar,\beta),
\eeq
and the quantum heat functional converges to the classical trajectory heat in the   weak coupling regime and $\hbar\rightarrow 0$ and $\beta\rightarrow 0$ limit. We note that for an Ohmic spectrum of the heat bath
\beq
J(\omega)=\pi\sum_{k}\frac{c_{k}^{2}}{2m\omega_{k}} \delta(\omega-\omega_{k})=\gamma \omega, \label{Ohmic}
\eeq
the classical bath correlation function satisfies $K(t)=2\gamma \delta(t)$ with $\gamma$ being the friction coefficient, and the classical trajectory heat reproduces the fluctuating heat defined by Sekimoto~\cite{Sekimoto}
\beq
Q_{\text{cl}}[X,\Omega]=-\frac{\gamma}{M^{2}}\int^{\tau}_{0}dt P^{2}(t)+\int^{\tau}_{0}dt \frac{P(t)}{M}\Omega(t), \label{CLQU}
\eeq
where $P(t):=M\dot{X}(t)$ is the classical momentum.


We next take the classical limit of Eq.~(\ref{CFH}). By taking $\hbar\rightarrow 0$, we can use the standard treatment for the stationary phase approximation. We also note that the quantum heat functional converges to the classical fluctuating heat~(\ref{supp:classicalq}). We introduce the Wigner function
\beq
p_{\rm{S}}(X_{i},P_{i}):=\int d\xi_{i} e^{\frac{i}{\hbar}P_{i}\xi_{i}}\rho_{\rm{S}}(X_{i},\xi_{i}), \label{Wigner}
\eeq
with $P_{i}:=M\dot{X}_{i}$, which converges to the corresponding classical phase space distribution in the $\hbar\rightarrow 0$ limit. Now Eq.~(\ref{CFH}) becomes (see Appendix~\ref{appendix:cal} for details)
\beqa
\chi_{Q}(\nu)&=&\int dX(0)dX(\tau) \int DX D\Omega P[\Omega] \delta( \mathcal{M}[X,\Omega])\nonumber \\
& &\times   p_{\rm{S}}(X_{i},P_{i})e^{i\nu Q_{\text{cl}}[X,\Omega]}+O(\hbar,\beta), \label{chiclassicala}
\eeqa
with
\beqa
\mathcal{M}[X,\Omega]&:=&M\ddot{X}(t)+V'[\lambda_{t},X(t)]+\int^{t}_{0}dsK(t-s)\dot{X}(s) \nonumber \\
& &+K(t)X_{i}-\Omega(t).
\eeqa
Here, $\mathcal{M}[X,\Omega]=0$ is the classical non-Markovian Langevin equation~(\ref{EQMLE}) for a product initial state~(\ref{supp:initial}). This implies that the delta function $\delta(\mathcal{M}[X,\Omega])$ in Eq.~(\ref{chiclassicala}) restricts the trajectory $X(t)$ of the system to a classical trajectory satisfying the Langevin equation. Therefore, from Eq.~(\ref{chiclassicala}), the classical limit of the characteristic function of heat converges to its classical counterpart, i.e.,
\beq
\chi_{Q}(\nu) = \langle e^{i\nu Q_{\text{cl}}}\rangle_{\text{cl-path}}+O(\hbar,\beta),
\eeq
where $\langle \bullet \rangle_{\text{cl-path}}$ means average over all classical paths. This completes the proof of the quantum-classical correspondence of the heat statistics.

We note that the quantum-classical correspondence of the work statistics has been shown in isolated~\cite{QCcorrespondence1,QCcorrespondence2} and open~\cite{Funo17} systems.

\section{\label{sec:summary}Conclusion}
We study the heat statistics for the Caldeira-Leggett model by using the path integral formulation of quantum mechanics. By integrating out the bath degrees of freedom, we introduce the heat functional [Eq.~(\ref{Qfunction})] along every pair of Feynman paths of the system. We show the microscopic reversibility between the heat functional and the ratio of the path probabilities [Eq.~(\ref{DBR})], which is useful to improve our understandings about heat in open quantum systems. We further show that the obtained heat functional reproduces the classical trajectory heat in the classical ($\hbar\rightarrow 0$ and $\beta\rightarrow 0$) limit. In addition, we show the quantum-classical correspondence of the heat statistics. This allows us to establish connections between the classical and quantum stochastic thermodynamics based on the Brownian motion model. It is left for future study for obtaining $\hbar$ corrections to the classical fluctuating heat. The obtained path integral formalism to study quantum heat in the Caldeira-Leggett model can handle non-Markovian and non rotating wave approximation regime and also for arbitrary time-variation of the Hamiltonian of the system. Therefore, we expect that the path integral approach discussed in this paper would contribute to future works about quantum thermodynamics in this interesting regimes.

\begin{acknowledgments}
This work was supported by the National Science Foundation of China under Grants No.~11775001 and 11534002, and The Recruitment
 Program of Global Youth Experts of China.
\end{acknowledgments}

\appendix

\section{\label{supp:sectwo}Derivation of the Feynman path based heat functional and the characteristic function of heat}
In this section, we derive the path-integral expression of the characteristic function of heat~(\ref{CFH}). Note that we can start from Eq.~(\ref{supp:EXHB}) and integrate out the bath degrees of freedom explicitly as done in Ref.~\cite{Carrega15}. However, we utilize a different technique used in Ref.~\cite{Diosi} to obtain Eq.~(\ref{CFH}). For convenience, let us introduce an operator
\beq
\rho_{\nu}(\tau):=e^{-\frac{i\nu}{2}H_{\rm{B}}}U_{\rm{SB}}(\rho_{\rm{S}}(0)\otimes \rho_{\rm{B}}^{\rm{G}}e^{i\nu H_{\rm{B}}} ) U^{\dagger}_{\rm{SB}}e^{-\frac{i\nu}{2}H_{\rm{B}}} \label{s:rhonu},
\eeq
which satisfies $\chi_{Q}(\nu)=\Tr[\rho_{\nu}(\tau)]$ and $\rho_{\nu}(0)=\rho_{\rm{S}}(0)\otimes \rho_{\rm{B}}^{\rm{G}}$. We move to the interaction picture $\rho_{\nu}^{\rm{I}}(t)=U^{\dagger}_{0}(t)\rho_{\nu}(t)U_{0}(t)$ with 
\beq
U_{0}(t):= \hat{\text{T}}\exp[-\frac{i}{\hbar}\int^{t}_{0}ds (H_{\rm{S}}(\lambda_{s})+\mu\hat{x}^{2})]\otimes e^{-\frac{i}{\hbar}H_{\rm{B}}t}.\label{s:unitary}
\eeq
Here, for technical simplicity, we include the counter term $\mu\hat{x}^{2}$ into the system Hamiltonian and define Eq.~(\ref{s:unitary}). Then, the evolution equation for $\rho_{\nu}^{\rm{I}}(t)$ takes the form
\beq
\frac{d}{dt}\rho_{\nu}^{\rm{I}}(t)=\frac{i}{\hbar}[\hat{x}^{\rm{I}}(t)\hat{B}^{\rm{I}}(t-\frac{\hbar\nu}{2})\rho_{\nu}^{\rm{I}}(t)-\rho_{\nu}^{\rm{I}}(t)\hat{x}^{\rm{I}}(t)\hat{B}^{\rm{I}}(t+\frac{\hbar\nu}{2})],\label{s:rhoder}
\eeq
where $\hat{B}^{\rm{I}}(t):=\sum_{k}c_{k}\hat{q}^{\rm{I}}_{k}(t)$. We integrate Eq.~(\ref{s:rhoder}) and obtain
\beqa
\chi_{Q}(\nu)&=&\Tr[\hat{T}\exp(\frac{i}{\hbar}\hat{\chi})(\rho_{\rm{S}}(0)\otimes\rho_{\rm{B}}^{\rm{G}})] \nonumber \\
&=&\Tr_{\rm{S}}[\hat{T}\exp(-\frac{1}{2\hbar^{2}}\Tr_{B}[\hat{T}\hat{\chi}^{2}\rho_{\rm{B}}^{\rm{G}}])], \label{s:chiQQ}
\eeqa
Here, we use the identity following Ref.~\cite{Diosi} and derive the second line, and $\hat{\chi}$ is defined by
\beq
\hat{\chi}:=\int^{\tau}_{0}dt \left[ \hat{x}^{\rm{I}}_{\rm{L}}(t)\hat{B}^{\rm{I}}_{\rm{L}}(t-\frac{\hbar\nu}{2})-\hat{x}^{\rm{I}}_{\rm{R}}(t)\hat{B}^{\rm{I}}_{\rm{R}}(t+\frac{\hbar\nu}{2})\right],
\eeq
where the subscripts $L$ and $R$ denote the operators acting on $\rho_{\nu}^{\rm{I}}(t)$ from the left and right, respectively~\cite{Diosi}. By using the bath correlation function $L(t)=\Tr_{\rm{B}}[\hat{B}^{\rm{I}}(t)\hat{B}^{\rm{I}}(0)\rho_{\rm{B}}^{\rm{G}}]$ given in Eq.~(\ref{bathcorr}), we obtain
\beqa
& &\Tr_{B}[\hat{T}\hat{\chi}^{2}\rho_{\rm{B}}^{\rm{G}}]=-2\int^{\tau}_{0}dt\int^{\tau}_{0}ds L(s-t+\hbar\nu)\hat{x}^{\rm{I}}_{\rm{L}}(t)\hat{x}^{\rm{I}}_{\rm{R}}(s) \nonumber \\
& &+2\int^{\tau}_{0}dt\int^{t}_{0}ds L(t-s)(\hat{x}^{\rm{I}}_{\rm{L}}(t)\hat{x}^{\rm{I}}_{\rm{L}}(s)+\hat{x}^{\rm{I}}_{\rm{R}}(t)\hat{x}^{\rm{I}}_{\rm{R}}(s)). \label{s:correll}
\eeqa
We combine Eqs.~(\ref{s:chiQQ}) and (\ref{s:correll}) and move to the path integral expression [note that $\hat{x}^{\rm{I}}_{\rm{L}}(t)\rightarrow x(t)$ and $\hat{x}^{\rm{I}}_{\rm{R}}(s)\rightarrow y(s)$], and finally obtain the characteristic function of heat~(\ref{CFH}).

Now let us briefly show that Eq.~(\ref{CFH}) agrees with the result obtained in Ref.~\cite{Carrega15}. Note that our notation is slightly different from the one used in Ref.~\cite{Carrega15}. In particular, we use the opposite sign convention for the definition of the heat $Q$ and thus $\nu$ as well. Let us introduce the following functions~\cite{Carrega15}
\beqa
L_{1}^{\nu}(t)&=&-\sum_{k}\frac{c_{k}^{2}}{2m_{k}\omega_{k}}\frac{\sin \frac{\nu\omega_{k}}{2}\sinh \frac{\omega_{k}(\beta-i\nu)}{2}}{\sinh\frac{\omega_{k}\beta}{2}}\cos\omega_{k}t \nonumber \\
\ L_{2}^{\nu}(t)&=&i\sum_{k}\frac{c_{k}^{2}}{2m_{k}\omega_{k}}\frac{\sin \frac{\nu\omega_{k}}{2}\cosh \frac{\omega_{k}(\beta-i\nu)}{2}}{\sinh\frac{\omega_{k}\beta}{2}}\sin\omega_{k}t. \nonumber
\eeqa
Then, the following relation holds:
\beq
2iL_{1}^{\nu}(t)+2iL_{2}^{\nu}(t)=L(t+\hbar\nu)-L(t). \label{s:LL}
\eeq
Now let us compare $i\Delta \Phi^{(\nu)}$ introduced in Ref.~\cite{Carrega15} and $i\nu Q_{\nu}$~[see Eq.~(\ref{Qfunction})] as follows:
\beqa
i\Delta \Phi^{(\nu)}&:=&i\int^{\tau}_{0}dt\int^{t}_{0}ds\left\{ \eta(t)\eta(s)-\xi(t)\xi(s)\right\}L_{1}^{\nu}(t-s) \nonumber \\
& &+i\int^{\tau}_{0}dt\int^{t}_{0}ds  \{\eta(t)\xi(s)-\xi(t)\eta(s)\}L_{2}^{\nu}(t-s)\nonumber \\
&=&2i\int^{\tau}_{0}dt\int^{\tau}_{0}ds(L^{\nu}_{1}(s-t)+L^{\nu}_{2}(s-t))x(t)y(s) \nonumber \\
&=&i\nu Q_{\nu}[x,y].
\eeqa
Here, $\eta(t)=x(t)+y(t)$ and $\xi(t)=x(t)-y(t)$, and we use Eq.~(\ref{s:LL}) and obtain the last equality. Therefore, the characteristic function of heat distribution~(\ref{CFH}) agrees with that obtained in Ref.~\cite{Carrega15}.

\section{\label{supp:secE}Some properties of the energy functional}
In this section, we discuss some properties of the energy functional introduced in Eq.~(\ref{initialEfunctional}).

First, the characteristic function of the initial energy distribution can be expressed as follows:
\beqa
\chi_{\epsilon}(E^{\rm{S}}(\lambda_{0}))&:=&\frac{1}{Z_{\rm{S}}(\lambda_{0})}\int dx_{i}dy_{i}\delta(x_{i}-y_{i})\int^{\bar{x}(\hbar\beta)=x_{i}}_{\bar{x}(0)=y_{i}} D\bar{x}_{0} \nonumber \\
& &\times e^{-\frac{1}{\hbar}S^{\mathrm{E}}[\bar{x}_{0},\lambda_{0}]+\epsilon E_{-i\epsilon}[\bar{x}_{0},\lambda_{0}]}, \label{initialGF}
\eeqa
where $E^{\rm{S}}(\lambda_0)$ is the energy of the system when the work parameter is equal to $\lambda_0$. Using Eq.~(\ref{initialGF}), we can obtain the n-th moment of the initial energy distribution as
\beq
\langle [E^{\rm{S}}(\lambda_{0})]^{n} \rangle=\partial_{\epsilon}^{n}\chi_{\epsilon}(E^{\rm{S}}(\lambda_{0}))|_{\epsilon=0}.
\eeq

Second, in the classical limit $(\hbar\rightarrow 0)$, the initial energy functional reproduces the classical internal energy of the system:
\beq
E_{\beta}[\bar{x}_{0},\lambda_{0}]=\frac{M}{2}[\dot{\bar{x}}(0)]^{2}+V[\bar{x}(0),\lambda_{0}]+O(\hbar),
\eeq
where $\dot{\bar{x}}(0)$ and $\bar{x}(0)$ are the classical velocity and position of the system at $t=0$, respectively.

\section{\label{appendix:cal}Some details of the classical limit of the characteristic function of the heat}
In this section, we derive Eq.~(\ref{heatdecomp}). We start by performing integration by parts in Eq.~(\ref{Qlimita}) and obtain
\beqa
 Q_{\nu}[x,y]
&=&\int^{\tau}_{0}dt \dot{X}(t)\Omega(t)-X(\tau)\Omega(\tau)+X(0)\Omega(0) \nonumber \\ 
& &- \int^{\tau}_{0}dt\int^{t}_{0} ds \dot{x}(t)\dot{y}(s)K(t-s) \nonumber \\
& &+\frac{1}{2}\Bigl( x_{i}y_{f}+y_{i}x_{f}\Bigr) K(t) -\frac{1}{2}\Bigl( x_{i}y_{i}+x_{f}y_{f}\Bigr) K(0) \nonumber \\
& &+\frac{1}{2} \int^{\tau}_{0}dt \Bigl( x_{f}\dot{y}(t)+y_{f}\dot{x}(t)\Bigr)K(\tau-t) \nonumber \\
& &- \frac{1}{2}\int^{\tau}_{0}dt \Bigl( x_{i}\dot{y}(t)+y_{i}\dot{x}(t)\Bigr) K(t) +O(\hbar). \label{supp:heatcala}
\eeqa
Here, we use the relation $\partial_{t}K(t)=2\text{Im}[L(t)]$ and derive Eq.~(\ref{supp:heatcala}). We note that $\xi(t)$ and $\dot{\xi}(t)$ describe quantum fluctuations from the classical coordinates $X(t)$ and $\dot{X}(t)$. Therefore, we shall neglect the terms proportional to $\xi(t)$ and $\dot{\xi}(t)$ in Eq.~(\ref{supp:heatcala}) in the classical limit. This procedure is essentially replacing $x(t)$ and $y(t)$ by $X(t)$ in Eq.~(\ref{supp:heatcala}). After some simplification, we obtain Eq.~(\ref{heatdecomp}).

Next, we give details of the derivation of Eq.~(\ref{chiclassicala}). We first expand the potential energy as $V(x)-V(y)=V(X+\xi/2)-V(X-\xi/2)=\xi V'(X)+O(\xi^{3})$ and obtain the lowest order terms of the forward and backward actions:
\beqa
\frac{i}{\hbar}S[x]-\frac{i}{\hbar}S[y]&=&-\frac{i}{\hbar}\int^{\tau}_{0}dt \xi(t)\Bigl( M\ddot{X}(t)+V'(X) \Bigr) \nonumber \\
& &-\frac{i}{\hbar}M\xi_{i}\dot{X}_{i}+ O(\xi^{3}) \label{supp:SXY}.
\eeqa
Next, we perform integration by parts in the Feynman-Vernon influence functional~(\ref{FV}) and obtain
\begin{widetext}
\beqa
F_{\text{FV}}[x,y]&=&\exp\Bigl[ -\frac{i}{\hbar}\int^{\tau}_{0}dt \xi(t)\Bigl( \int^{t}_{0}ds K(t-s)\dot{X}(s)+K(t)X_{i}\Bigr)-\frac{1}{\hbar}\int^{\tau}_{0}dt\int^{\tau}_{0}ds \text{Re}[L(t-s)]\xi(t)\xi(s) \Bigr] \nonumber \\
&=&\int D\Omega P[\Omega] \exp\Bigl[ -\frac{i}{\hbar}\int^{\tau}_{0}dt \xi(t)\Bigl( \int^{t}_{0}ds K(t-s)\dot{X}(s)+K(t)X_{i}-\Omega(t)\Bigr)\Bigr]. \label{supp:FVT}
\eeqa
\end{widetext}
Here, we use the Habburd-Stratonovich transformation and introduce the noise function~(\ref{supp:NOISE}) and the weight function~(\ref{weightF}). Combining Eqs.~(\ref{supp:SXY}) and (\ref{supp:FVT}), the classical limit of the characteristic function of heat is given by
\begin{widetext}
\beqa
\chi_{Q}(\nu)&=&\int dX_{f}dX_{i}\int d\xi_{i} \int DX \int D\xi \int D\Omega P[\Omega]  e^{-\frac{i}{\hbar}M\dot{X}_{i}\xi_{i}}\rho(X_{i},\xi_{i})e^{i\nu Q_{\text{cl}}[X]} \nonumber \\
& &\times \exp\biggl[ -\frac{i}{\hbar}\int^{\tau}_{0}dt \xi(t)\Bigl( M\ddot{X}(t)+V'[X(t)] +\int^{t}_{0}ds K(t-s)\dot{X}(s)+K(t)X_{i}-\Omega(t)\Bigr) \biggr]+O(\hbar,\beta), \label{supp:sysfinal}
\eeqa
\end{widetext}
Here, we assume the weak coupling regime and use the classical limit of the heat functional~(\ref{CLQU}) in Eq.~(\ref{supp:sysfinal}). We finally obtain Eq.~(\ref{chiclassicala}) by performing the integral $\int D\xi$ [which gives $\delta(\mathcal{M}[X,\Omega])$] and introducing the Wigner function~(\ref{Wigner}).

\section{\label{supp:secone}Derivations of the classical trajectory heat in the classical Brownian motion model}
In what follows, we derive the classical heat in the classical Brownian motion model. The obtained expression is used to show the quantum-classical correspondence of the heat functional and its statistics.


\subsection{Derivation of the underdamped Langevin equation from the Hamiltonian of the composite system}
We start from the composite system modeled by the classical Brownian motion model and derive the underdamped Langevin equation, which describes the reduced dynamics of the system~\cite{Zwanzig,Lindenberg}. This technique is utilized to relate the energy change of the heat bath with the classical heat in stochastic thermodynamics.

The Hamiltonian of the classical Brownian motion model is given by Eq.~(\ref{Hamiltonian}), where we replace all operators by c-numbers, i.e., $\hat{x}\rightarrow X$, $\hat{p}\rightarrow P$, $\hat{x}_{k}\rightarrow q_{k}$, $\hat{p}_{k}\rightarrow p_{k}$, etc.
We assume that the initial probability distribution of the composite system is given by the following product state
\beq
\rho(0)=\rho_{\rm{S}}(X(0),P(0))\frac{1}{Z_{\rm{B}}(0)}\exp\Bigl[ -\beta H_{\rm{B}}(q_{k}(0),p_{k}(0))\Bigr], \label{supp:initial}
\eeq
where $\rho_{\rm{S}}(X(0),P(0))$ is an arbitrary distribution of the system. The Hamilton equations for the composite system can be written as
\beqa
\dot{X}&=&\frac{\partial H}{\partial P}=\frac{P}{M},  \\
\dot{P}&=&-\frac{\partial H}{\partial X}=-\frac{\partial V(\lambda_{t},X)}{\partial X}+\sum_{k}\Bigl( c_{k}q_{k}-\frac{c^{2}_{k}X}{m_{k}\omega_{k}^{2}}\Bigr) ,\hspace{5mm} \label{EQMP}\\
\dot{q}_{k}&=& \frac{\partial H}{\partial p_{k}}=\frac{p_{k}}{m_{k}} ,\\
\dot{p}_{k}&=&-\frac{\partial H}{\partial q_{k}}=-m_{k}\omega_{k}^{2}q_{k}+c_{k}X.
\eeqa
Now we can formally solve the equation of motion as follows:
\beqa
q_{k}(t)&=&q_{k}(0)\cos\omega_{k}t+\frac{p_{k}(0)}{m_{k}\omega_{k}}\sin\omega_{k}t \nonumber \\
& &+\frac{c_{k}}{m_{k}\omega_{k}}\int^{t}_{0}ds\sin\omega_{k}(t-s)X(s), \label{EQMq}\\
p_{k}(t)&=&-m_{k}\omega_{k}q_{k}(0)\sin\omega_{k}t+p_{k}(0)\cos\omega_{k}t \nonumber \\
& &+c_{k}\int^{t}_{0}ds\cos\omega_{k}(t-s)X(s).\label{EQMp}
\eeqa
Substituting Eq.~(\ref{EQMq}) into Eq.~(\ref{EQMP}) yields the non-Markovian Langevin equation
\beqa
& &\dot{P}(t)+\frac{\partial}{\partial X}V(\lambda_{t},X)+\frac{1}{M}\int^{t}_{0}ds K(t-s)P(s) \nonumber \\
& &+K(t)X(0)=\Omega(t). \label{EQMLE}
\eeqa
The term $K(t)X(0)$ is referred to as the initial slippage term which describes a fast relaxation of the bath to the canonical distribution conditioned on the initial state $X(0)$ of the system. Here,
\beq
\Omega(t)=\sum_{k}c_{k}\left(q_{k}(0)\cos\omega_{k}t +\frac{p_{k}(0)}{m_{k}\omega_{k}}\sin\omega_{k}t \right),
\eeq
is the noise and $K(t-s)$ defined in Eq.~(\ref{clbathcorre}) is the classical bath correlation function. We interpret the initial random preparation of the bath coordinates and momenta as the source of the noise. Therefore, the noise average is defined by
\beq
\av{f[\Omega(t)]}=\int\prod_{k}dq_{k}(0)dp_{k}(0)\frac{\exp(-\beta H_{\rm{B}}(0))}{Z_{B}}f[\Omega(t)],
\eeq
where $f[\Omega(t)]$ is an arbitrary function of the noise. We can show that the noise satisfies the following properties:
\beqa
\av{\Omega(t)}&=&0 \\
\av{\Omega(t)\Omega(s)}&=&\beta^{-1}K(t-s).
\eeqa

Next, we consider a Markovian dynamics of the system by taking the Ohmic spectrum~(\ref{Ohmic}). Then, the noise becomes the Gaussian white noise
\beq
\av{\Omega(t)\Omega(s)}=2kT\gamma\delta(t-s).
\eeq
Now the equation of motion~(\ref{EQMLE}) becomes the Markovian Langevin equation with inertia terms
\beq
\dot{P}(t)+\frac{\partial}{\partial X}V(\lambda_{t},X)+\frac{\gamma}{M} P(t)+2\gamma\delta(t)X(0)=\Omega(t). \label{supp:MarkovLE}
\eeq

\subsection{Derivation of the fluctuating heat}
We now derive the expression for the fluctuating heat from the energy change of the heat bath plus the interaction. Using Eqs.~(\ref{EQMq}) and (\ref{EQMp}), the change of the interaction energy can be expressed as
\beqa
\Delta H_{\rm{SB}}&=&-\sum_{k}c_{k}\left( X(\tau)q_{k}(\tau)-X(0)q_{k}(0)\right)\nonumber \\
& &+\sum_{k}\frac{c_{k}^{2}}{2m_{k}\omega_{k}^{2}}(X^{2}(\tau)-X^{2}(0)) \nonumber \\
&=&- \frac{1}{2}\Bigl(X^{2}(\tau)-X^{2}(0)\Bigr)K(0)-X(\tau)X(0)K(0)\nonumber \\
& &-X(\tau)\Omega(\tau)+X(0)\Omega(0) \nonumber \\
& &+X(\tau)\int^{\tau}_{0}dt K(\tau-t)\frac{P(t)}{M}. \label{supp:deltaSB}
\eeqa
Also, the energy change of the heat bath takes the form
\beqa
\Delta H_{\rm{B}} &=&\sum_{k}\left( \frac{p_{k}^{2}(\tau)-p_{k}^{2}(0)}{2m_{k}}+\frac{m_{k}\omega_{k}^{2}}{2}(q_{k}^{2}(\tau)-q_{k}^{2}(0))\right) \nonumber \\
&=&-\Delta H_{\rm{SB}} +\frac{1}{M^{2}}\int^{\tau}_{0}dt \int^{t}_{0}ds  P(t)P(s)K(t-s)\nonumber \\
& & -\int^{\tau}_{0}dt \frac{P(t)}{M}\Omega(t) +X(0)\int^{\tau}_{0}dt K(t)\frac{P(t)}{M}. \label{supp:deltaHB}
\eeqa

We now define heat by the change in the energy of the heat bath $Q:=-\Delta H_{\rm{B}}$. Because we can neglect the interaction energy in the weak coupling regime,  we have
\beqa
Q&=&-\frac{1}{M^{2}}\int^{\tau}_{0}dt \int^{t}_{0}ds  P(t)P(s)K(t-s) \nonumber \\
& &+\int^{\tau}_{0}dt \frac{P(t)}{M}\Omega(t)-X(0)\int^{\tau}_{0}dt K(t)\frac{P(t)}{M}. \hspace{3mm} \label{supp:classicalQ}
\eeqa
The last term in Eq.~(\ref{supp:classicalQ}) is related to the initial slippage term $K(t)X(0)$ in the Langevin equation~(\ref{EQMLE}), and it describes the heat generated by the fast relaxation of the heat bath to the conditional canonical distribution. By assuming the Ohmic spectrum~(\ref{Ohmic}), we reproduce the stochastic heat for the classical Markovian dynamics of the system:
\beq
Q=-\frac{\gamma}{M^{2}}\int^{\tau}_{0}dt P^{2}(t)+\int^{\tau}_{0}dt \frac{P(t)}{M}\Omega(t)-\frac{\gamma}{M} X(0)P(0). \label{supp:CLQU}
\eeq

We finally discuss the heat related to the initial slippage term $Q_{\text{slip}}$~(\ref{slip}). Suppose we consider the initial state
\beq
\rho'(0)=\rho_{\rm{S}}(X(0),P(0))\frac{1}{Z_{B}}\exp\Bigl[ -\beta (H_{\rm{B}}+H_{\rm{SB}})\Bigr] \label{supp:condinitial}
\eeq
instead of Eq.~(\ref{supp:initial}). Then, the Langevin equation~(\ref{supp:MarkovLE}) now takes the form
\beq
\dot{P}(t)+\frac{\partial}{\partial X}V(\lambda_{t},X)+\frac{\gamma}{M} P(t) =\Omega(t). \label{supp:MarkovLEA}
\eeq
This equation of motion is usually used in the study of classical stochastic thermodynamics, and the fluctuating heat $Q=-\Delta H_{\rm{B}}$ reproduces the definition of the stochastic heat defined by Sekimoto~\cite{Sekimoto}:
\beq
Q=-\frac{\gamma}{M^{2}}\int^{\tau}_{0}dt P^{2}(t)+\int^{\tau}_{0}dt \frac{P(t)}{M}\Omega(t). \label{supp:CLQC}
\eeq
Because we can approximate Eq.~(\ref{supp:condinitial}) by Eq.~(\ref{supp:initial}) in the weak coupling regime, it is reasonable to neglect the initial slippage term $2\gamma\delta(t)X(0)$ in the Langevin equation~(\ref{supp:MarkovLE}). Therefore, we can neglect the heat $Q_{\text{slip}}$~(\ref{slip}) associated with the fast relaxation of the system described by the initial slippage term in the weak coupling regime.

\end{document}